\documentclass[12pt,preprint]{aastex}

\newcommand{\etal  }{{et al.} }
\newcommand{\msun}{\thinspace M_\odot}  
\newcommand{\rsun}{\thinspace R_\odot}  
\newcommand{\vect}[1]{\mbox{\boldmath$#1$}}

\def\lesssim{\mathrel{\hbox{\rlap{\hbox{\lower4pt\hbox{$\sim$}}}\hbox{$<$}}}}
\def\gtrsim{\mathrel{\hbox{\rlap{\hbox{\lower4pt\hbox{$\sim$}}}\hbox{$>$}}}}
\newcommand{\cm  }{\,{\rm cm}^{-3} } 
\newcommand{\nc  }{n_{\rm c} }

\newcommand{\dfrac}[2]{{\displaystyle \frac{#1}{#2}}  }

\shorttitle{The First Jet in the Universe}
\shortauthors{Machida  \etal 2006}

\begin{document}

\title{The First Jet in the Universe: Protostellar Jets from the First Stars}

\author{Masahiro N. Machida\altaffilmark{1} , Kazuyuki Omukai\altaffilmark{2}, Tomoaki Matsumoto\altaffilmark{3}, and Shu-ichiro Inutsuka\altaffilmark{1}} 

\altaffiltext{1}{Department of Physics, Graduate School of Science, Kyoto University, Sakyo-ku, Kyoto 606-8502, Japan; machidam@scphys.kyoto-u.ac.jp, inutsuka@tap.scphys.kyoto-u.ac.jp}
\altaffiltext{2}{National Astronomical Observatory of Japan, Mitaka, Tokyo 181-8588, Japan; omukai@th.nao.ac.jp}
\altaffiltext{3}{Faculty of Humanity and Environment, Hosei University, Fujimi, Chiyoda-ku, Tokyo 102-8160, Japan; matsu@i.hosei.ac.jp}

\begin{abstract}
The protostellar jets driven by the formation of the first stars are studied by using three-dimensional magnetohydrodynamical (MHD) nested grid simulations.
Starting from a slowly rotating spherical cloud of $5.1 \times 10^4 \msun$ permeated by a uniform magnetic field, we follow the evolution from the central number density $\nc = 10^3 \cm$ 
(where the radius of the object $r = 6.6$ pc) to $n_c \simeq 10^{23} \cm$ ( $r \simeq 1 \rsun$).
We resolve the cloud structure more than 8 orders of magnitude in spatial extent 
and 20 orders in density contrast. 
We calculate four models that differ in initial magnetic field strengths and angular velocities.
In all models, protostars of $\simeq 10^{-3}\msun$ are formed at $\nc \simeq 10^{22} \cm$ in accordance with one-dimensional calculations.
By this epoch, the magnetic flux density is amplified by 10 orders of magnitude from the initial value.
Consequently, the formed protostar possesses the magnetic field of $\sim10^6$ G that is much larger than the flux density of the present counterparts, reflecting the fact that the dissipation of a magnetic field is ineffective in primordial gas clouds.
If the initial magnetic field $B > 10^{-9} (\nc/10^3 \cm)^{2/3}\,{\rm G}$, the protostellar jet is launched  and its velocities reaches $\sim 70$ km s$^{-1}$ by the time the protostellar mass becomes $(4-6)\times 10^{-3} \msun$, and a fraction ($3-10\%$) of the accreting matter is blown off from the central region.
Owing to the interaction of these ejecta with surrounding matter, expanding bow shocks are created at both heads of the jet.
If this jet continues to sweep out the surrounding gas that otherwise accretes onto the central star or circumstellar disk, the final mass of the first star can be substantially reduced.
In addition, dense post-shock regions behind the bow shocks are expected to promote the chemical reactions (formation of H$_2$ and HD), and this provides possible environments for subsequent low-mass star formation in the early universe.
\end{abstract}

\section{Introduction}
Magnetic field, which is an indispensable ingredient in the present-day local interstellar medium (ISM), has been largely overlooked in studies of the first star formation. 
This is probably because the magnetic force has been believed to be unimportant in comparison with either the gravity or the thermal pressure in the early universe.

% Theoretical study of the collapsing primordial gas shows that  pressure $P$ in the cooling gas scales with density $\rho$  as $P \propto \rho^{\gamma_{\rm eff}}$ where  $\gamma_{\rm eff} \sim 1.1$ (e.g., Omukai \& Nishi 1998). 
Theoretical studies showed that a collapsing primordial gas follows a
self-similar solution with pressure $P$ in the cooling gas scaling with
density $\rho$ as $P \propto \rho^{\gamma_{\rm eff}}$ where $\gamma_{\rm eff} \sim 1.1$ \citep[e.g.,][]{omukai98}.
 In a gravitationally collapsing state such as  in self-similar solutions, the relative importance of pressure and gravity inside of  the runaway collapsing core remains almost the same. 
In contrast,  the magnetic field in the homologously collapsing medium scales as $B \sim \rho^{2/3}$,  and thus, the magnetic pressure increases more rapidly than the gas pressure. 
In general, if the magnetic field scales as $B \sim \rho^{\gamma_B}$ and $\gamma_B > \gamma_{\rm eff}$,  the dynamical importance of the magnetic field increases  continuously in the collapsing  primordial cloud. 
Therefore we cannot entirely neglect the effect of the magnetic  field in the collapsing primordial gas even if it is relatively  small initially. 
\citet{ichiki06} showed that magnetic fields are created by the cosmological fluctuations when they enter the cosmic horizon before the cosmic recombination epoch. 
By this mechanism, the magnetic field of $\sim 10^{-15}-10^{-14}$G is produced at the scale of the first cosmological objects ($\sim 10^{5-6}M_{\sun}$).
\citet{langer03} proposed another mechanism for magnetic field generation during the epoch of the cosmic reionization.
They found that magnetic fields in the intergalactic matter on the order of $\sim 10^{-11}\ {\rm G}$ are created  at the scale of protogalaxies, which increase up to $\simeq 10^{-7}-10^{-8}\ {\rm G}$ if the objects experience the spherical collapse up to number density $n = 10^3 \cm$. These fields can influence the evolution of primordial gas.

In this paper, we study the evolution of weakly magnetized primordial star-forming cores and the formation of protostars using three-dimensional numerical magnetohydrodynamical (MHD) simulations.
As a result, we demonstrate that protostellar jets are driven by the formed protostars even in the primordial condition.

\section{Model and Numerical Method}
 To study the evolution of primordial star-forming cores taking account of a magnetic field and self-gravity in a large dynamic range of density and spatial scale, we use the three-dimensional ideal MHD nested grid method.
The ideal MHD equations including the self-gravity are  
\begin{eqnarray} 
& \dfrac{\partial \rho}{\partial t}  + \nabla \cdot (\rho \vect{v}) = 0, & \\
& \rho \dfrac{\partial \vect{v}}{\partial t} 
    + \rho(\vect{v} \cdot \nabla)\vect{v} =
    - \nabla P - \dfrac{1}{4 \pi} \vect{B} \times (\nabla \times \vect{B})
    - \rho \nabla \phi, & \\ 
& \dfrac{\partial \vect{B}}{\partial t} = 
   \nabla \times (\vect{v} \times \vect{B}), & \\
& \nabla^2 \phi = 4 \pi G \rho, &
\end{eqnarray}
 where $\rho$, $\vect{v}$, $P$, $\vect{B} $, and $\phi$ denote the density, velocity, pressure, magnetic flux density, and gravitational potential, respectively.
The ideal MHD approximation is justified in our case.
Studying the coupling of the magnetic field in star-forming cores, \citet {maki04} concluded that magnetic field is frozen to the gas for the initial field strength below $10^{-5} 
(n_{\rm H}/10^3 \cm)^{0.55} {\rm G}$.
For gas pressure, we use a barotropic relation that approximates the result of one-zone calculation where thermal and chemical processes for primordial gas are solved in detail \citep{omukai05}.
This relation (thick solid line) as well as the original data (thick dotted line) for pressure evolution is plotted as a function of number density in the upper panel 
of Figure~\ref{fig:1}. 
To stress variations of pressure with density, we plot $P/n$, which is proportional to the gas temperature if the mean molecular weight is constant.

As the initial states, we take a spherical cloud whose density is twice higher than that for hydrostatic equilibrium with external pressure
\citep[i.e., so called Bonnor-Ebert sphere,][]{bonnor56,ebert55}.
The cloud rotates rigidly ($\Omega_0$) around the $z$-axis and is threaded by a uniform magnetic field $B_0$ parallel to the rotation axis.
The initial central density is taken as $ \nc = 10^3 \cm $, which corresponds to the epoch of formation of star-forming cores \citep{abel00,bromm99,bromm02}.
The radius and mass inside the sphere are $R_{\rm c} = 6.6$ pc and $ M_{\rm c} = 5.1 \times 10^4 \msun $, respectively.
The initial temperature is 250 K. 
After the formation of the star-forming core, self-gravity of gas dominates the dark matter gravity (Abel et al. 2002).
Thus, we do not include the dark matter component in our calculation.
The models are characterized by two parameters: initial magnetic field strength ($B_0$) and initial angular velocity ($\Omega_0$).
We calculate four models denoted as B6, B7, B8 and B9, whose model parameters are ($B_0$, $\Omega_0$) =  ($10^{-6}$ G, $3.3 \times 10^{-16}$ s$^{-1}$; Model B6), ($10^{-7}$ G, $3.3 \times 10^{-17}$ s$^{-1}$; Model B7),  ($10^{-8}$ G, $2.1 \times 10^{-17}$ s$^{-1}$; Model B8),  and ($10^{-9}$ G, $1.3 \times 10^{-17}$ s$^{-1}$; Model B9), respectively.
We have observed fragmentation of clouds in cases with faster rotation. 
For example, with initial magnetic field $B_0 = 10^{-6}\ {\rm G}$, the cloud fragments if angular velocity $\Omega_0 > 3.3 \times 10^{-16}$ s$^{-1}$.
Our rotation rates are close to the critical values for fragmentation according to our numerical experiment.
Since our focus in this paper is on driving of protostellar jet rather than on core fragmentation, we adopt somewhat slow rotation rates in comparison with some examples of \citet{bromm02} who studied the fragmentation of primordial clouds by hydrodynamical simulation. 
We leave the stage of the evolution with larger rotation rate to our subsequent papers.

We adopt the nested grid method \citep[for detail, see ][]{machida05a,machida05b} to achieve high spatial resolution near the origin.
Each level of a rectangular grid has the same number of cells ($ = 64 \times 64 \times 32 $), although the cell width $h(l)$ depends on the grid level $l$.
The cell width is reduced by a factor of two for every upper level.
The calculation is started with three grid levels ($l=1,2,3$).
Box size of the coarsest grid $l=1$ is chosen to be $2 R_{\rm c}$. 
At the boundary, the magnetic field and ambient gas are assumed to rotate with the constant angular velocity $\Omega_0$ \citep[for detail see][]{matsumoto04}.
A new finer grid is generated whenever the minimum local Jeans length $ \lambda _{\rm J} $ becomes smaller than $ 8\, h (l_{\rm max}) $, where $h$ is the cell width.
The maximum level of grids is restricted to $l_{\rm max} \leqq 30$.
Since the density is highest in the finest grid, generation of new grid ensures the Jeans condition of \citet{truelove97} with a margin of a safety factor of 2.

\section{Results}
\begin{figure}
\plotone{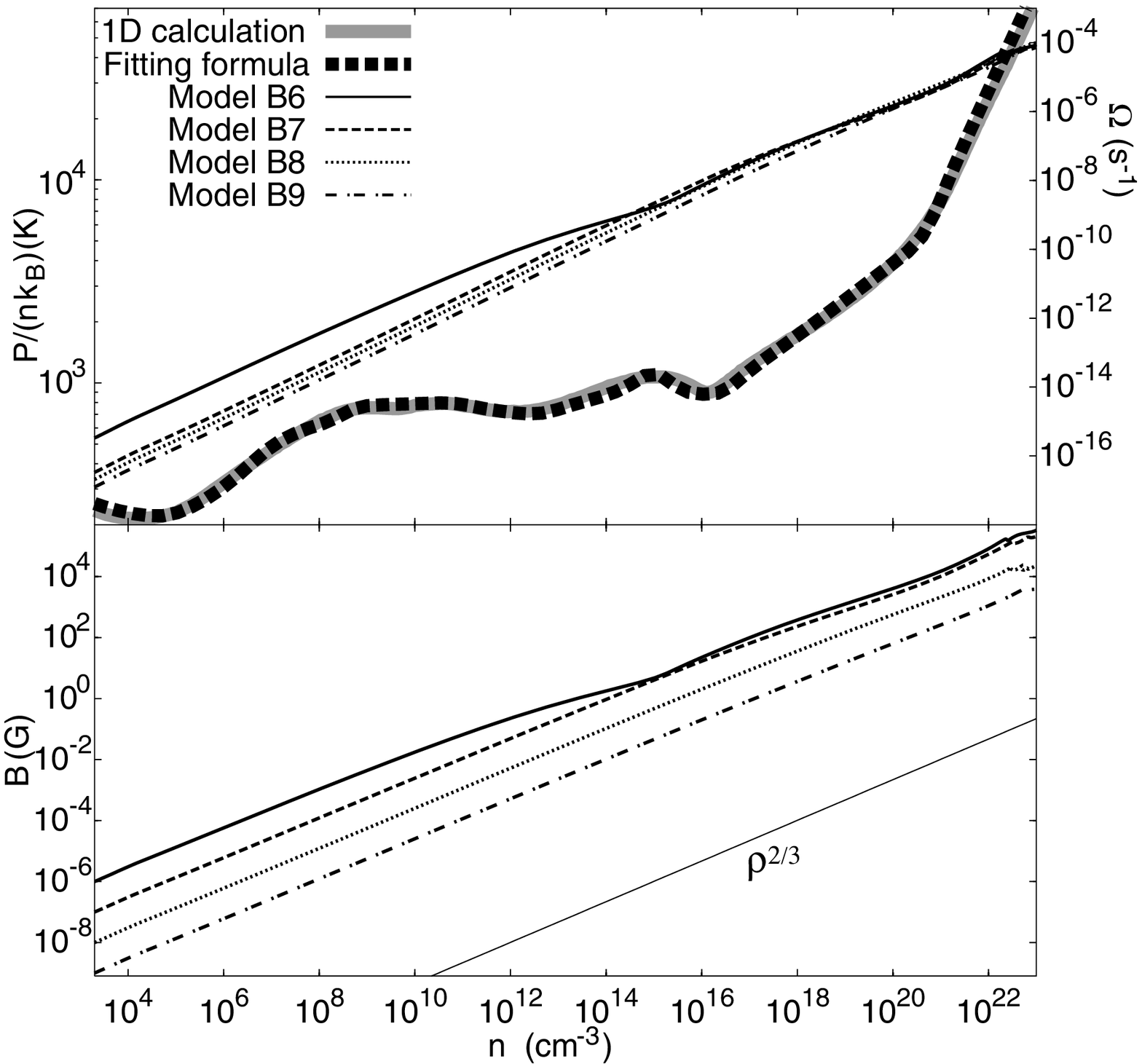}
\caption{
Upper panel: The gas pressure normalized by number density derived from a one-dimensional calculation (thick solid line) and our fit used in the numerical simulation (thick dotted line) are plotted.
The central angular velocities ($\Omega_{\rm c}$) against the number density ($\nc$) are also plotted for models B6 (solid line), B7 (broken line), B8 (dotted line), and B9 (dash-dotted line).
Lower panel:  The magnetic flux densities ($B_{\rm c}$) against the central density for models B6 (solid line), B7 (broken line), B8 (dotted line), and B9 (dash-dotted line).
The relation $\propto \rho^{2/3}$, valid for the spherical collapse,  
is also plotted for comparison.
}
\label{fig:1}
\end{figure}

 Starting from the number density $\nc = 10^3 \cm$, the evolution of primordial clouds is calculated up to $ \simeq 10^{23} \cm$.
 Figure~\ref{fig:1} shows the evolution of central angular velocities ($\Omega_{\rm c}$; upper panel) and magnetic flux densities ($B_{\rm c}$; lower panel) against the central number density ($\nc$) for all models.  
Both of angular velocity and magnetic flux density in model B6 increase in proportion to $ \nc^{2/3}$, which is the case of spherical collapse, for $10^3\cm < \nc \lesssim 10^{14}\cm$ \citep[see][]{machida05a,machida06a}.  
For higher densities, the growth rates in unit of logarithmic density ($d \log B_{\rm c}/d \log \nc$ and $d \log \Omega_{\rm c}/d \log \nc$) become smaller, and  the angular velocity and magnetic flux density increase in proportion to $\nc^{1/2}$, because cloud collapses non-homologously into the self-similar disk-like structure \citep{machida05a,machida06a}.
On the other hand, in models B7, B8, and B9, both the magnetic flux densities and angular velocities continue to increase in proportion to $\nc^{2/3}$ until $\nc\simeq 10^{21} \cm$.
This indicates that, owing to small magnetic field and slow rotation, disk formation is delayed and spherical collapse continues until very high density.
In all models, the angular velocities have converged to the same scaling law before $\nc \simeq 10^{19}\cm$ owing to the change of geometry of the collapse and magnetic braking: for stronger magnetic field, the magnetic braking is more effective and the growth rate of angular velocity tends to be smaller.
Since we assume rapid (slowly) rotation for strongly (weakly, respectively) magnetized clouds in this paper, the angular velocities tend to converge to the unique scaling law as the collapse proceeds.

At $\nc \simeq 10^{20}\cm$ that corresponds to the completion of molecular hydrogen dissociation, the equation of state becomes adiabatic as shown in the upper panel of Figure~\ref{fig:1}.
The hydrostatic core of mass $M_{\ast} \simeq 10^{-3} \msun$ is formed at $\nc \simeq 10^{22}\cm$ in all models in agreement with one-dimensional calculation \citep{omukai98}.
We call this hydrostatic core a protostar hereafter.
The protostar rotates at angular velocity of $\Omega \simeq 10^{-4}$ s$^{-1}$ ($\simeq 8.63$ day$^{-1}$) initially, which is comparable to or slightly faster than those of the present-day pre-main sequence stars  \citep{herbst02}.
By this time, the magnetic field has been amplified to the level of $B_{c} \simeq 6 \times 10^5$ G (Model B9) $-$ $6.1\times 10^6$ G (Model  B6) that is much stronger than that of the present day protostar or, so called, the second core \citep[see][]{machida06b}.
This is because the magnetic field in primordial gas does not dissipate either by Ohmic dissipation or ambipolar diffusion, as shown by \citet{maki04}.
The formed protostars in the present-day star formation process have the magnetic field of only $\sim 10^3$ G,  owing to effective Ohmic dissipation \citep{machida06b}.

\begin{figure}
\plotone{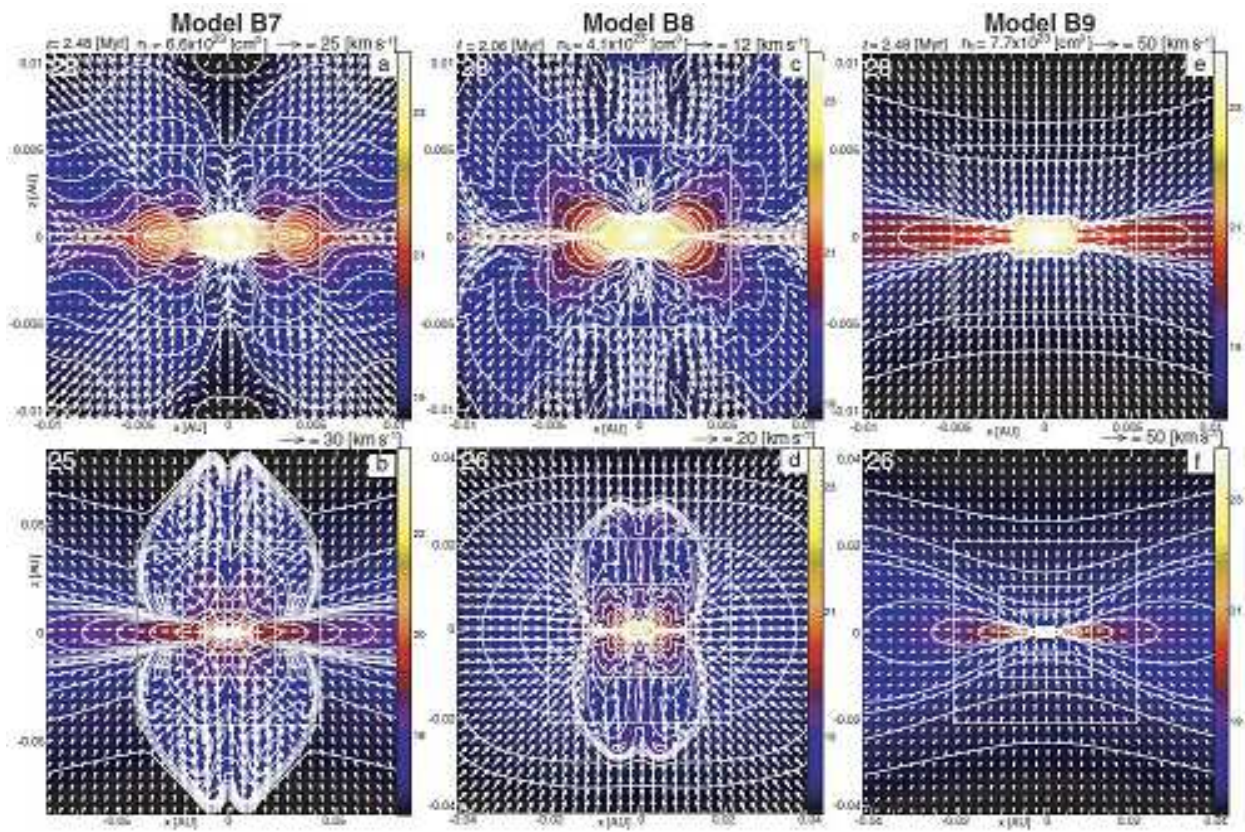}
\caption{
The density (color-scale and white contours) and velocity distribution (arrows) on the cross-section in the $y=0$ plane for models B7 (panels [a] and [b]), B8 (panels [c] and [d]), and B9 (panels [e] and [f]).
Model name, elapsed time, central number density, and velocity scale are plotted in each upper panel.
Grid level ($l$) are displayed at upper left corner.
Each upper panel is a close-up view of the corresponding lower panel. 
}
\label{fig:2}
\end{figure}

Figure~\ref{fig:2} shows the density distributions and velocity vectors on the $y=0$ cut planes for models B7, B8, and B9 at the end of the calculation.
Each upper panel is a close-up view of the corresponding lower panel.
The level of the outermost grid is indicated at the upper left corner of each panel.
Finer grids are overplotted by boxes in each panel.
We omit model B6 from this figure, since the evolution of model B6 is almost the same as that of model B7.
We can see in Figure~\ref{fig:2}{\it a} that a strong jet is driven from the central protostar.

 A spherical protostar of radius $0.65 \rsun$ forms at the center, which is surrounded by a disk-like structure extending up to $\simeq 2 \rsun$.
 The jet starts flowing 49.1 hours after the protostar formation epoch.\footnote{ We define the protostar formation epoch as the time when the central density reaches $n_c = 10^{20} \cm$, where the equation of state becomes adiabatic.}
 Rippling of density contours in Figure~\ref{fig:2} indicate the occurrence of strong mass ejection near the center.
The mass of outflowing gas is $7\times 10^{-4}\msun$, while the mass of the protostar and the disk are $4.4 \times 10^{-3} \msun$ and $1.6 \times 10^{-3}\msun$, respectively, at the end of the calculation.
Namely, about 10\% of the accreted matter is ejected as the jet.
The bow shocks are generated at the top of the jet, $z\simeq \pm 0.1$ AU (Fig.~\ref{fig:2}{\it b}), owing to interaction with ambient matter.
The jet is well-collimated and creates a cocoon-like structure as shown in Figure~\ref{fig:2}{\it b}.
At this epoch, the maximum speed of the jet is 66.3 km s$^{-1}$.
In model B6, the mass accretion rate, outflow rate, the speed of the jet, and density structure are very similar to those in model B7.

In model B8, the jet appears 45 hours after the protostar formation epoch.
The speed of the jet in model B8 is slower than that in model B7.
We can see from the arrows in Figure~\ref{fig:2}{\it c} and {\it d} that the speeds of jet are comparable to or slower than the infall speed in model B8, while the speeds of the jet in model B7 are much larger than the infall speed (Fig.~\ref{fig:2}{\it a} and {\it b}).
 However, the rippling density contours indicative of strong mass ejection are also seen in this model (Fig.~\ref{fig:2}{\it c}).
 The mass of the core, disk, and outflowing gas increase up to $6.15\times 10^{-3}\msun$, $1.25 \times 10^{-3} \msun$, and $2.5 \times 10^{-4} \msun$, respectively, by the end of the calculation.
 Thus, about 3\% of the accreted matter is converted to outflow until the end of the calculation in this model.
Owing to smaller magnetic field, the outflow efficiency is reduced by about a factor of three from model B7.

Further reduction of magnetic field results in the absence of jet in model B9.
In this model, an oblate protostar forms inside a thin disk as shown in Figure~\ref{fig:2}{\it e} and {\it f}. 
Although we followed the evolution of the accreting protostar up to 201 hours after the protostar formation epoch, we could not observe any sign of the jet.
Thus, without hindrance by jet, the gas can accrete onto the protostar directly in the vertical direction as well as through the disk, while, in other models (B6-8), the gas accretes mostly through the thin disk.
After the central protostar forms, magnetic field inside the star is significantly amplified both by the magneto-rotational instability \citep{balbus91} and by the shearing motion between the star and the ambient medium.
Consequently, the plasma beta $\beta_{\rm p}$, which is the ratio of gas pressure to magnetic pressure, inside the star becomes almost unity, i.e. the magnetic energy is comparable to the thermal energy.
Just outside the disk, however, the magnetic energy is much smaller than the thermal energy ($\beta_{\rm p} \gg 1$).
This small magnetic energy compared with the thermal energy is the reason why the jet is not launched in model B9.
In contrast, in models B6-8, the magnetic energy is comparable to or lager than the thermal energy even outside the disk.
Note, however, that a long time calculation of model B9 remain to be carried out to clarify whether a jet is eventually launched or not.

\section{Conclusion and Discussion}
From these experiments, we conclude that the protostellar jet is driven by the primordial protostar if the initial magnetic field $B \ga 10^{-9}$\,G at $n = 10^3 \cm$.
This critical value corresponds to the magnetic field of $5\times 10^{-13}$ G in the background material ($\nc \simeq 0.01 \cm$) if the spherical collapse ($B \propto \rho^{2/3}$) is assumed. 
Whether this initial magnetic field is realized or not remains uncertain.
The magnetic field expected from the cosmological fluctuations falls below this value about an order of magnitude \citep{takahashi05,ichiki06}.
However, with slightly more amplification by an unknown mechanism, the field could exceed the critical value.
The field created at the cosmic reionization ($\sim 10^{-11}$G) seems to be sufficient (Langer et al. 2003).
However, after the reionization epoch, the primordial star formation occurs only in large halos with the virial temperature exceeding $10^{4}$ K, being strongly suppressed in smaller halos by UV background radiation \citep[e.g.,][]{kitayama01}.
Even if the initial magnetic field is weaker than our critical value, the protostellar jet could be launched in a later phase of star formation.
By analytically studying the evolution of accretion disks around the first stars, \citet{tan04} suggested that magnetic fields amplified in the disk eventually give rise to protostellar outflows during the protostellar accretion phase.
Since we did not follow the evolution up to the late phase considered by \citet{tan04}, we cannot draw any conclusions concerning their expectation.

Note that we assumed rather slow rotation of clouds ($\Omega_0 = 3.3\times 10^{-16} {\rm s^{-1}} - 1.3\times 10^{-17} {\rm s^{-1}}$) as the initial state.
Strictly speaking, this critical value for the magnetic field strength is only valid for clouds having small angular velocities.
When an initial cloud rotates rapidly, the jet might not be driven by the protostar because the amplification rate of magnetic field  changes according to the geometry of the collapse and cloud fragmentation \citep{machida05a,machida06a}.
This possibility will be studied in our subsequent works.

Although numerical studies on fragmentation of first cosmological objects concluded that fragmentation mass scale of metal-free gas is as massive as $100-1000 \msun$ \citep[e.g.,][]{bromm99,bromm02,abel00,omukai03}, they are limited to the case of star formation from the cosmological initial condition, which is unaffected by astrophysical phenomena.
Jet and outflow from the first stars disturb the condition of star forming regions in the early universe.
What kinds of stars are formed in such regions? 
Observations of local star-forming regions indicate that the protostellar outflows compress surrounding matter and induce subsequent star formation \citep[e.g.,][]{bally99,quillen05}.
Since our calculations demonstrated that the jets are indeed driven by the first stars, similar events may possibly occur also in the case of the first stars.

When the velocity of jets exceeds $\sim 50$ km s$^{-1}$, gas is heated above $\sim10^5$ K and then ionized.
If the shocked layer becomes fully ionized,  a large amount of molecular hydrogen forms in the postshock flow by using electron as a catalyst during slow recombination \citep{maclow86,shapiro87,susa98}.
The enhanced H$_2$ cooling realizes low-temperature environment ($\la 150$K) favorable for HD formation, and this HD cooling further reduces the temperature below $\sim 100$ K.
In such cold clumps, low-mass stars are expected to be produced \citep[e.g.,][]{uehara00,nakamura02,machida05c,nagakura05}.
Those low-mass stars survive to the present and could be observed in the halo.
\citet{suda04} indicated that the extremely metal-poor ([Fe/H]$<$-5) stars \citep{christlieb01,frebel05} discovered recently were formed as binary members from metal-free gas, and then have been polluted by companion stars during the stellar evolution.

However, there is a caveat in discussing the subsequent star formation around the first stars.
UV photons emitted by the formed massive star may dissociate H$_2$ and HD molecules in the surrounding matter thereby suppressing the subsequent star formation (Omukai \& Nishi 1999).
Future study on induced star formation by protostellar outflow should include this effect.

Not only the subsequent star formation but also the first star formation process itself is affected by the outflow. 
For example, in the present-day star formation scenario, the final stellar mass is envisaged to be affected by the mass outflow (for low-mass stars) or radiative feedback (for massive stars) from the protostar.
These mechanisms stop mass accretion from the parent cloud to the protostar, as well as blow off the parent cloud itself \citep{nakano95}.
In the case of primordial star formation, the radiative feedback is insufficient for halting the accretion of metal-free gas because of the lack of dust grain (Omukai \& Palla 2001; 2003).
The mass of first stars grows up to $\sim 600 \msun$ if the mass accretion continues unimpeded.
The outflow from the first star may reduce its final stellar mass by blowing off a considerable fraction of the cloud matter before accreting onto the star.
At the same time, the jet may reduce the star formation efficiency by diverting material from the accretion disk.

The jets from the first stars also contribute to disperse the magnetic field lines into the intergalactic medium.
This may affect the formation of next generation stars and galaxies.
To study this effect further, we need to calculate the evolution  up to the later phase possibly by adopting implicit numerical integration method for MHD equations.

\acknowledgments
We greatly benefited from discussion with T. Kato, K. Saigo, and H. Susa.
We also thank T. Hanawa for contribution in developing the nested grid code.
Numerical calculations were carried out with a Fujitsu VPP5000 at the Astronomical Data Analysis Center, the National Astronomical Observatory of Japan.
This work was supported partially by the Grants-in-Aid from MEXT (15740118, 16077202, 16740115).

\clearpage

\end{document}